\documentclass[aps,prl,reprint,groupedaddress]{revtex4-2}

\usepackage{graphicx,amsmath,physics,xcolor,amssymb} 
\usepackage{tikz}

\usetikzlibrary{quantikz2}

\usepackage{dcolumn}
\usepackage{bm}

\usepackage[utf8]{inputenc}
\usepackage[T1]{fontenc}
\usepackage{mathptmx}
\usepackage{etoolbox}

\begin{document}


\title{A Quantum Algorithm for Simulating Nonunitary Dynamics Governed by Nonautonomous Linear Ordinary Differential Equations}


\author{Pouya Khazaei}
\affiliation{Department of Chemistry, University of Michigan, Ann Arbor, MI  48109, U.S.A.}

\author{Eitan Geva}
\email[]{eitan@umich.edu}

\affiliation{Department of Chemistry, University of Michigan, Ann Arbor, MI  48109, U.S.A.}


\date{\today}

\begin{abstract}
Nonautonomous linear ordinary differential equations of the form $\dot{v}(t) = A(t)\, v(t)$, where $A(t)$ is non-skew-symmetric, are often used to describe nonunitary dynamics in a variety of fields that range from open quantum system dynamics to economic modeling.  
Because quantum computing hardware is designed to natively implement unitary transformations, existing algorithms for solving such equations on quantum hardware are based on the assumption that the nonunitary propagator is known, and use dilation techniques to embed the nonunitary dynamics within the unitary dynamics of a larger system. However, with the exception of cases where the nonunitary propagator is known in closed form, it needs to be calculated and manipulated on a classical computer at each time step. 
In this paper, we propose a quantum algorithm that does not require a priori knowledge of the explicit nonunitary propagator and effectively performs the dilation on the quantum hardware. 
Our algorithm combines a dilation scheme that uses singular value decomposition (SVD) to write the nonunitary propagator as a sum of unitaries with simulating the dynamics of the SVD factors on the quantum hardware.  
The population-only time-convolutionless quantum master equation describing photoinduced charge transfer in a solvated molecular triad is used as a demonstrative example of the applicability of the algorithm and its sensitivity to noise. 
%
%
\end{abstract}


\maketitle


\section{Introduction}

Nonautonomous linear ordinary differential equations (ODEs) of the form $\dot{v}(t) = A(t)\, v(t)$, where $v(t)$ is a vector and $A(t)$ is a matrix,
are used as equations of motion (EOMs) 
in many research fields that range from chemistry, biology and physics to engineering and economics \cite{kato_hierarchical_2019,chaturvedi_time-convolutionless_1979,shamma_gain_1992,leith_survey_2000,jin_reconstruction_2018,kloeden_nonautonomous_2013}. 
While such equations can include complex-valued $A(t)$ and $v(t)$, 
complex-valued linear ODEs can always be written in terms of real-valued ones by
decomposing them into coupled ODEs for the real and imaginary parts. 
Likewise, many partial differential equations (PDEs) reduce to sets of coupled linear nonautonomous ODEs upon spatial discretization. 
Thus, an algorithm for solving real nonautonomous linear ODEs can be straightforwardly extended to complex nonautonomous linear ODEs and PDEs. We will therefore focus on cases where $A(t)$ and $v(t)$ are real. 

Often times, the generator of the dynamics described by nonautonomous linear ODEs, $A(t)$, is {\em non-skew-symmetric} (i.e. $A^T(t) \neq -A(t)$, where $A^T(t)$ is the transpose of $A(t)$), 
which leads to {\em nonunitary} dynamics. At the same time, 
quantum hardware is designed to natively implement {\em unitary} circuits \cite{nielsen_quantum_2010}. 
Thus, simulating nonunitary dynamics on quantum hardware often involves dilation 
into a larger Hilbert space in which the overall evolution is unitary \cite{stinespring_positive_1955,nielsen_quantum_2010}.
The nonunitary propagator is then realized as a subblock of a higher-dimensional unitary matrix and recovered through measurement and post-selection \cite{hu_quantum_2020}. The Sz.-Nagy construction \cite{sz-nagy_harmonic_2010} provides one general dilation strategy, though at the cost of increased Hilbert-space dimension and circuit depth, and has been explored in chemically motivated simulations \cite{lyu_simulating_2024,wang_simulating_2023}. More recently, singular value decomposition (SVD)-based dilation schemes have been proposed \cite{schlimgen_quantum_2022,dan_simulating_2025}, which are based on factorizing the propagator into orthogonal matrices and a diagonal matrix of singular values, often reducing circuit depth and improving suitability for noisy intermediate-scale quantum (NISQ) implementations.

Importantly, existing SVD-based dilation algorithms are inherently hybrid quantum-classical. Specifically, unless it is known in closed form, the nonunitary propagator needs to be computed and SVD-ed at each time step, which is a task performed on a classical computer. 
For an $N$-dimensional system, this procedure requires classical propagation of an $N \times N$ matrix and repeated $\mathcal{O}(N^3)$ SVD computations, effectively treating each time slice independently and discarding the smooth geometric structure of the underlying matrix flow. As a result, the quantum computer ends up executing a sequence of precomputed operators, such that the dominant computational cost remains classical.

In this paper, we present a 
quantum algorithm for implementing the aforementioned SVD-based dilation that avoids the need for a classically precomputed nonunitary propagator and its SVD.  Specifically, instead of repeatedly computing and dilating the propagator at each time step on a classical computer, we derive coupled EOMs for the SVD factors of the time-dependent propagator and evolve them too on the quantum computer. 

\section{Theory}
\label{sec:theory}

\subsection{Linear Nonautonomous Ordinary Differential Equations}

A linear ODE 
whose coefficient matrix depends explicitly on time is called a \emph{nonautonomous} dynamical system \cite{strogatz_nonlinear_2015}. 
In the real-valued case, the equation takes the form
\begin{align}
    \dot{v}(t) = A(t)\, v(t)~~, 
    \label{eq:gen_qme}
\end{align}
where $v(t) \in \mathbb{R}^N$ denotes the state vector at time $t$, and $A(t) \in \mathbb{R}^{N \times N}$ is a time-dependent coefficient matrix. 
When $A$ is {\em time-independent}, the system is said to be \emph{autonomous}. 
When $A$ is explicitly {\em time-dependent}, the system is said to be \emph{nonautonomous}.

Many EOMs of interest can be written in this form. For example, a vectorized time-convolutionless quantum master equation (tcl-QME) reduces to Eq.~\eqref{eq:gen_qme} \cite{shibata_generalized_1977,chaturvedi_time-convolutionless_1979,lai_simulating_2021,trushechkin19}.
In contrast, commonly used Markovian quantum master equations such as the Redfield or Lindblad equations \cite{redfield57,nitzan06,mukamel_principles_1995,alicki87,davies74}
typically involve time-independent generators and therefore correspond to autonomous linear ODEs. The algorithm developed here applies generally to nonautonomous linear ODEs and is not restricted to quantum dynamical problems.

Let $\Phi(t)$ denote the propagator, also known as {\em the principal fundamental matrix solution} of Eq.~\eqref{eq:gen_qme}, such that 
\begin{align}
    v(t) = \Phi(t)\,v(0).
    \label{eq:Phi}
\end{align}
Substituting Eq. \ref{eq:Phi} into Eq. \ref{eq:gen_qme}, and noting that
Eq. \ref{eq:Phi} must be satisfied for any $v(0)$, implies that $\Phi(t)$ must also satisfy Eq. \ref{eq:gen_qme}:
\begin{align}
    \dot{\Phi}(t) = A(t)\,\Phi(t), 
    \qquad 
    \Phi(0) = I,
\end{align}
where $I$ is the identity matrix. $\Phi(t)$ is unitary when $A(t)$ is skew-symmetric ($A^T = -A$) and nonunitary when $A(t)$ is non-skew-symmetric ($A^T \neq -A$). In what follows we will focus on the case of nonunitary $\Phi(t)$. 



\subsection{Quantum Algorithm for the System Dynamics}

In the case of a nonunitary $\Phi(t)$,
implementation on a quantum computer requires a dilation procedure.
To this end, we adopt a dilation procedure based on singular-value decomposition (SVD) which was proposed in Ref. \citenum{schlimgen_quantum_2022}. 
Within this dilation method, SVD needs to be performed on $\Phi(t)$ at each and every time step:
\begin{align}
    \Phi(t) = U(t)\,\Sigma(t)\,V^{\dagger}(t)~~.
    \label{eq:svd}
\end{align}
Here, $U(t)$ and $V(t)$ are unitary matrices, while $\Sigma(t)$ is a nonunitary diagonal matrix:
\begin{align}
    \Sigma(t)
    &= \mathrm{diag}\big(\sigma_{1}(t), \ldots, \sigma_{N}(t)\big)
 \notag \\  &= \left( 
    \begin{array}{ccc}
\sigma_1(t) & \cdots & 0 \\
\vdots & \ddots & \vdots \\
0 & \cdots & \sigma_N (t)
    \end{array}
    \right)
    ~~,
\end{align}
where $\{ \sigma_j(t) | j=1,\ldots,N\}$ are the real and non-negative {\em singular values}.

In the next step, we order the singular values by magnitude such that $\sigma_1 (t)$ is the largest:
\begin{align}
    \sigma_{1}(t)
    \ge \sigma_{2}(t)
    \ge \cdots
    \ge \sigma_{N}(t)~~,
\end{align}
and rescale them relative to the largest singular value:
\begin{align}
    \tilde{\sigma}_{j}(t)
    = \frac{\sigma_{j}(t)}{\sigma_{1}(t)}~~,~~
    \qquad
    0 \le \tilde{\sigma}_{j}(t) \le 1.
    \label{sigtilde}
\end{align}
The diagonal matrix $\Sigma(t)$ can then be written as a sum of two unitaries:
\begin{align}
    \Sigma(t)
    = \frac{\sigma_{1}(t)}{2}
      \left(\Sigma^{+}(t) + \Sigma^{-}(t)\right),
      \label{sigma_pm}
\end{align}
where $\Sigma^{\pm}(t)$ are diagonal unitary matrices whose diagonal elements are given by:
\begin{align}
    \bigl(\Sigma^{\pm}(t)\bigr)_{jj}
    =
    \tilde{\sigma}_{j}(t)
    \,\pm\,
    i\sqrt{1 - \tilde{\sigma}_{j}^{2}(t)}.
    \label{eq:diag_spm}
\end{align}
Since  $| \tilde{\sigma}_{j}(t) | \in [0,1]$, the quantities
\(
\tilde{\sigma}_{j}(t)
\pm
i\sqrt{1-\tilde{\sigma}_{j}^{2}(t)}
\)
lie on the unit circle, ensuring that
$\Sigma^{\pm}(t)$ are unitary.

The state vector is assumed to be normalized, such that:
\begin{align}
    \| v(0) \| = 1,
\end{align}
If $v(0)$ is not normalized,
it must be rescaled so that it is normalized.

\begin{figure}
    \centering
    \begin{quantikz}
        \lstick{$\ket{v(0)}$} & \gate{V^{\dagger}} & \gate[wires=2]{U_{\Sigma}}
                                & \gate{U} & \meter{} \\
        \lstick{$\ket{0}$}       & \gate{H}          &
                                                        & \gate{H} & \meter{}
    \end{quantikz}
    \caption{Quantum circuit implementing the SVD-based dilation using
    \(V^{\dagger}\), \(U_{\Sigma}\), and \(U\) on the system register and
    Hadamard gates on the ancilla.}
    \label{fig:dilation-circuit}
\end{figure}
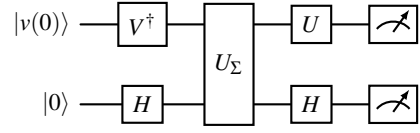

The full time evolution governed by Eq.~\eqref{eq:gen_qme}
is implemented by introducing a single ancilla qubit
and applying the circuit shown in Fig.~\ref{fig:dilation-circuit}.
Here, $U_{\Sigma}(t)$ is the following block-diagonal unitary acting on the joint
system–ancilla Hilbert space:
\begin{align}
    U_{\Sigma}(t)
    =
    \Sigma^{+}(t)
    \oplus
    \Sigma^{-}(t)~~,
\end{align}
and $H$ is the Hadamard gate, which prepares and recombines the ancilla superposition
which is required for the two dilation branches to interfere.

Acting on the initial joint state
$\ket{v(0)} \otimes \ket{0}$,
the circuit produces
\begin{align}
    \frac{1}{2}
    \begin{pmatrix}
        U(t)\left(\Sigma^{+}(t)+\Sigma^{-}(t)\right)
        V^{\dagger}(t)\ket{v(0)} \\
        U(t)\left(\Sigma^{+}(t)-\Sigma^{-}(t)\right)
        V^{\dagger}(t)\ket{v(0)}
    \end{pmatrix}.
\end{align}
Post-selecting the ancilla outcome $\ket{0}$ yields
\begin{align}
    \frac{\Phi(t)\ket{v(0)}}{\sigma_{1}(t)}.
\end{align}
Thus, successful realization of the target evolution corresponds
to measurement of the ancilla in the $\ket{0}$ state.

\subsection{Equations of motion for the SVD factors}


The dynamics of $\Phi(t)$ can be cast in terms of the dynamics of $U(t)$, $V(t)$ and $\Sigma (t)$ [see Eq.~\eqref{eq:svd}]. 
The EOMs for $U(t)$, $V(t)$ and $\Sigma (t)$ 
were derived in Ref.~\citenum{wright_differential_1992} and are given by:
\begin{align}
    \dot{U}(t) &= U(t)\, Z(t),
    \label{eq:udot} \\
    \dot{V}(t) &= V(t)\, W(t),
    \label{eq:vdot} \\
    \dot{\sigma}_i(t) &= G_{ii}(t)\, \sigma_i(t),
    \label{eq:sdot}
\end{align}
where $\{ G_{ii}(t) \}$ are the diagonal elements of the matrix
\begin{align}
    G(t) = U^{T}(t)\, A(t)\, U(t)~~,
\end{align}
and $Z(t)$ and $W(t)$ are matrices whose matrix elements are defined as follows:
\begin{align}
    Z_{jk}(t)
    &= \frac{\sigma_k^2(t)\, G_{jk}(t)
            + \sigma_j^2(t)\, G_{kj}(t)}
           {\sigma_k^2(t) - \sigma_j^2(t)}, \\
    W_{jk}(t)
    &= \frac{\sigma_k(t)\sigma_j(t)
            \left(G_{kj}(t) + G_{jk}(t)\right)}
           {\sigma_k^2(t) - \sigma_j^2(t)}.
\end{align}
It should be noted that Eqs.~\eqref{eq:udot}-\eqref{eq:sdot} have been studied extensively in dynamical systems theory,
including applications to the computation of Lyapunov exponents and
exponential dichotomy spectra \cite{dieci_svd_2008,dieci_lyapunov_2002,dieci_singular_2006,dieci_lyapunov_2007}. 
In this paper, we use them to develop a quantum algorithm for solving nonautonomous linear ODEs.

Importantly, $U(t)$ and $V(t)$ must remain orthogonal at all times:
\begin{align}
    U^{T}(t)U(t) = I,
    \qquad
    V^{T}(t)V(t) = I~~.
\end{align}
This is indeed guaranteed by the fact that the generators $Z(t)$ and $W(t)$ in Eqs.~\eqref{eq:udot} and \eqref{eq:vdot} are real skew-symmetric matrices:
\begin{align}
    Z^{T}(t) = -Z(t),
    \qquad
    W^{T}(t) = -W(t)~~.
\end{align}
This in turn means that the propagators of $U(t)$ and $V(t)$ correspond to orthogonal, or unitary, transformations. 
Thus, Eqs.~\eqref{eq:udot} and \eqref{eq:vdot} can be integrated directly
on a quantum computer using Lie-group integrators, such as
Magnus-expansion-based methods, 
which generate orthogonal transformations by construction.

In contrast, $\Sigma(t)$ is not orthogonal. However, given Eq.~\eqref{sigma_pm}, $\Sigma(t)$ can be cast in terms of $\Sigma^{\pm}(t)$, which are orthogonal by construction.
In the next step, we derive the corresponding EOMs
for $\Sigma^{\pm}(t)$.

\subsection{Equations of motion for $\Sigma^\pm (t)$}

The EOMs for 
$\Sigma^{\pm}(t)$ can be obtained by differentiating their
diagonal elements defined in Eq.~\eqref{eq:diag_spm}:
%
\begin{align}
    \dot{\sigma}^{\pm}_{j}(t)
    =
    \left(
        1
        \mp
        \frac{i\,\tilde{\sigma}_{j}(t)}
        {\sqrt{1 - \tilde{\sigma}_{j}^{2}(t)}}
    \right)
    \dot{\tilde{\sigma}}_{j}(t).
    \label{eq:spmdot1}
\end{align}
Recalling that $\tilde{\sigma}_{j}(t) = {\sigma_{j}(t)}/{\sigma_{1}(t)}$ (see Eq. \ref{sigtilde}), we find that
\begin{align}
    \dot{\tilde{\sigma}}_{j}(t)
    =
    \frac{\dot{\sigma}_{j}(t)}{\sigma_{1}(t)}
    -
    \frac{\dot{\sigma}_{1}(t)\,\sigma_{j}(t)}
         {\sigma_{1}^{2}(t)}.
\end{align}
Using Eq.~\eqref{eq:sdot}, $    \dot{\sigma}_{i}(t)
    =
    G_{ii}(t)\,\sigma_{i}(t),
$ we obtain
\begin{align}
    \dot{\tilde{\sigma}}_{j}(t)
    =
    \big(G_{jj}(t) - G_{11}(t)\big)
    \tilde{\sigma}_{j}(t).
    \label{eq:sigtildedot}
\end{align}

Substituting Eq.~\eqref{eq:sigtildedot}
into Eq.~\eqref{eq:spmdot1} then gives the equation of motion for the
elements of $\Sigma^{\pm}(t)$:
\begin{align}
    \dot{\sigma}^{\pm}_{j}(t)
    =
    \big(G_{jj}(t) - G_{11}(t)\big)
    \tilde{\sigma}_{j}(t)
    \left(
        1
        \mp
        \frac{i\,\tilde{\sigma}_{j}(t)}
        {\sqrt{1 - \tilde{\sigma}_{j}^{2}(t)}}
    \right).
\end{align}
This equation can be written in matrix form with a
time-dependent generator:
\begin{align}
    \dot{\Sigma}^{\pm}(t)
    =
    -\,i\,L^{\pm}(t)\,\Sigma^{\pm}(t),
\end{align}
where $L^{\pm}(t)$ are diagonal with elements given by
\begin{align}
    L^{\pm}_{jj}(t)
    =
    \pm
    \big(G_{jj}(t) - G_{11}(t)\big)
    \frac{\tilde{\sigma}_{j}(t)}
         {\sqrt{1 - \tilde{\sigma}_{j}^{2}(t)}}.
\end{align}

For $j=1$, we have $\tilde{\sigma}_{1}(t)=1$, which implies
\(
\sigma^{\pm}_{1}(t)=1
\)
for all times. Consequently,
\begin{align}
    L^{\pm}_{11}(t)=0.
\end{align}
Thus, all factors ${U(t), V(t), \Sigma^{\pm}(t)}$ evolve under generators that produce orthogonal transformations, enabling the full SVD-based dynamics to be implemented using unitary quantum gates.


\subsection{Summary of the workflow for solving nonautonomous linear ODE on a quantum computer}

We now summarize the 
workflow used to evolve the real nonautonomous linear ODE
\(
\dot{v}(t) = A(t)v(t)
\)
on a quantum computer. In what follows, we refer to this work flow and the algorithm it gives rise to as  \emph{q-SVD-EOM}.

The orthogonal factors
\(
\{U(t), V(t), \Sigma^{\pm}(t)\}
\)
evolve under generators that produce unitary (orthogonal) propagators
and can therefore be implemented directly on a quantum device.
The corresponding EOMs are
\begin{align}
    \dot{U}(t) &= U(t)\, Z(t),
    \label{eq:udot2} \\
    \dot{V}(t) &= V(t)\, W(t),
    \label{eq:vdot2} \\
    \dot{\Sigma}^{\pm}(t) &= -\, i\, L^{\pm}(t)\, \Sigma^{\pm}(t),
    \label{eq:sigpmdot}
\end{align}
where
\begin{align}
    G(t) = U^{T}(t)\, A(t)\, U(t).
\end{align}

The matrix elements of the generators are
\begin{align}
    Z_{jk}(t)
    &= \frac{\tilde{\sigma}_k^{2}(t)\, G_{jk}(t)
            + \tilde{\sigma}_j^{2}(t)\, G_{kj}(t)}
           {\tilde{\sigma}_k^{2}(t) - \tilde{\sigma}_j^{2}(t)}, \\
    W_{jk}(t)
    &= \frac{\tilde{\sigma}_k(t)\tilde{\sigma}_j(t)
            \left(G_{kj}(t) + G_{jk}(t)\right)}
           {\tilde{\sigma}_k^{2}(t) - \tilde{\sigma}_j^{2}(t)}, \\
    L^\pm_{jj}(t)
    &=
    \begin{cases}
        0, & j = 1, \\[6pt]
        \displaystyle
        \pm
        \frac{\tilde{\sigma}_j(t)
              \big(G_{jj}(t) - G_{11}(t)\big)}
             {\sqrt{1 - \tilde{\sigma}_j^{2}(t)}},
        & j \neq 1.
    \end{cases}
\end{align}

The above-mentioned steps in the workflow are given in terms of unitary transformations and can therefore be executed on quantum computer. 
The only remaining step in the workflow which is nonunitary, and therefore needs to be executed on a classical computer, is the dynamics of the largest singular value, $\sigma_1 (t)$, which is governed by the following scalar ODE:
\begin{align}
    \dot{\sigma}_{1}(t)
    = G_{11}(t)\, \sigma_{1}(t).
    \label{eq:s1dot}
\end{align}
Within, q-SVD-EOM, solving Eq. \ref{eq:s1dot} is the only computational task executed on a classical computer.  

\subsection{Quantum Algorithm for propagating $U$ and $V$}

\label{subsec:UV_algorithm}

The treatment of $\{U(t),V(t) \}$ differs from that of $\Sigma^\pm (t)$ because
$U$ and $V$ are real-valued orthogonal matrices whose generators are
skew-symmetric. In what follows we outline the algorithm for propagating $U (t)$ (the exact same algorithm is applied for propagating $V (t)$).

The EOM for $U(t)$ is $\dot{U}(t) = U(t)\, Z(t)$ [see Eq.~\eqref{eq:udot2}],
where $Z(t)$ is skew-symmetric.
Propagation from time $t_i$ to time $t_i+h$, where $h$ is the time step, is implemented via the Cayley transform \cite{iserles_cayley-transform_2001,korol_cayley_2019},
\begin{align}
    U(t_{i+1})
    =
    U(t_i)\, \operatorname{cay}(Z(t_i)),
\end{align}
where
\begin{align}
    \operatorname{cay}(Z)
    =
    \left[I - \tfrac{h}{2} Z\!\left(t_i+\tfrac{h}{2}\right)\right]^{-1}
    \left[I + \tfrac{h}{2} Z\!\left(t_i+\tfrac{h}{2}\right)\right].
\end{align}
For a skew-symmetric $Z$, $\operatorname{cay}(Z)$ is orthogonal and
therefore unitary.
Because $Z(t)$ depends on $U(t)$ and $\sigma(t)$, the midpoint value
is approximated using the midpoint extrapolation method (MPEA) of Ref.~\citenum{leitenmaier_heterogeneous_2022},
\begin{align}
    Z\!\left(t_i+\tfrac{h}{2}\right)
    =
    \tfrac{23}{12} Z(t_i)
    - \tfrac{16}{12} Z(t_{i-1})
    + \tfrac{5}{12} Z(t_{i-2}).
\end{align}

Rather than vectorizing $U$, we exploit the structure of right
multiplication. Writing $U$ in terms of its row vectors
$\{\mathbf{u}_i\}_{i=1}^N$,
\begin{align}
    U =
    \begin{pmatrix}
    \mathbf{u}_1 \\
    \mathbf{u}_2 \\
    \vdots \\
    \mathbf{u}_N
    \end{pmatrix},
\end{align}
the update becomes
\begin{align}
    \mathbf{u}_i(t_{i+1})
    =
    \operatorname{cay}(Z(t_i))^{T}\, \mathbf{u}_i(t_i).
\end{align}
Each row $\mathbf{u}_i \in \mathbb{R}^N$ is therefore propagated
independently, under the same linear transformation. Treating
$\mathbf{u}_i$ as a normalized quantum state,
\begin{align}
    |\mathbf{u}_i\rangle
    =
    \mathbf{u}_i,
\end{align}
the update is implemented as
\begin{align}
    |\mathbf{u}_i\rangle
    \;\longmapsto\;
    \operatorname{cay}(Z(t_i))^{T}\, |\mathbf{u}_i\rangle.
\end{align}
This reduces the matrix evolution to $N$ independent vector evolutions,
each of dimension $N$.
In contrast, a column-wise decomposition is not consistent with the
dynamics. Under right multiplication, each updated column is a linear
combination of all previous columns, and therefore does not evolve
independently.

Measurement yields the absolute squares of the components of
$\mathbf{u}_i(t_{i+1})$. Since $U$ is real and orthogonal, the
components are real up to a sign. Assuming no sign changes over a
sufficiently small time step, the updated rows are reconstructed using
the measured magnitudes together with the signs from the previous step.
Full quantum state tomography is therefore unnecessary.

Due to device noise, the reconstructed matrix $\tilde{U}(t_{i+1})$ is
generally not exactly orthogonal. To restore the required structure,
we enforce skew-symmetry at the generator level,
\begin{align}
    Z^{\text{skew}}(t_{i+1})
    =
    \tfrac{1}{2}
    \big[
    \tilde{Z}(t_{i+1})
    -
    \tilde{Z}(t_{i+1})^{T}
    \big],
\end{align}
where $\tilde{Z}$ is constructed from the noisy SVD factors.

If an orthogonal matrix is required explicitly, we project onto the
nearest orthogonal matrix via $    U^{\mathrm{proj}}(t_{i+1})
    =
    U_u V_u^{T},
$ where $U_u$ and $V_u$ are obtained from the SVD
of $\tilde{U}(t_{i+1})$. Alternative projection schemes, such as polar
decomposition or Gram--Schmidt orthogonalization, may also be employed.

\subsection{Quantum Algorithm for $\Sigma^\pm$}
\label{subsec:spm_algorithm}

The treatment of $\Sigma^\pm$ differs from that of $\{U,V\}$ because
$\Sigma^\pm$ are complex-valued diagonal matrices whose generators are
also diagonal. Since $\Sigma^- = \Sigma^{+*}$, it is sufficient to evolve
$\Sigma^+$.

Rather than vectorizing the full $N \times N$ matrix, we exploit its
diagonal structure and encode only its diagonal elements:
\begin{align}
    |\Sigma^+\rangle
    =
    \mathrm{diag}(\Sigma^+).
\end{align}
This reduces the Hilbert-space dimension from $N^2$ to $N$.

Because each diagonal element of $\Sigma^+$ has unit magnitude,
\begin{align}
    \langle \Sigma^+ | \Sigma^+ \rangle = N,
\end{align}
the normalized encoding is obtained by dividing by $\sqrt{N}$.

The vectorized equation of motion is
\begin{align}
    |\dot{\Sigma}^+\rangle
    =
    - i L^+ |\Sigma^+\rangle,
\end{align}
where $L^+$ is diagonal. Time evolution over a step
$t_i \rightarrow t_i+h$ is implemented via the Cayley transform,
\begin{align}
    |\Sigma_{i+1}^{+}\rangle
    =
    \operatorname{cay}(- iL^+) |\Sigma_i^{+}\rangle.
\end{align}
Because the generator is diagonal, the Cayley update does not mix
computational basis states. The normalized state therefore has the following form:
\begin{align}
    |\Sigma^+\rangle
    =
    \frac{1}{\sqrt{N}}
    \sum_{j=1}^{N}
    e^{i\phi_j} |j\rangle,
\end{align}
Thus, the evolution affects only the phases $\{\phi_j\}$. Since
$\tilde{\sigma}_1 = 1$, the first diagonal element 
correspond to $\phi_1 = 0$, so that the remaining phases are defined
relative to it.

The reconstruction problem thus reduces to determining the
relative phases $\{\phi_j\}_{j=2}^N$.
For each $j \ge 2$, we perform an interferometric measurement between
the reference component $|1\rangle$ and $|j\rangle$.
This is achieved by applying a Hadamard-type rotation in the
two-dimensional subspace spanned by $\{|1\rangle, |j\rangle\}$,
which mixes the two amplitudes and converts their relative phase
into measurable probability differences.
To access the remaining phase information, we first apply
a phase rotation (e.g., $S^\dagger$) to $|j\rangle$ before the
Hadamard mixing, which shifts the relative phase so that a second
measurement reveals the complementary information.
Repeating this procedure for each $j$ yields all relative phases.

Because the magnitudes of the amplitudes are known a priori and only
$N-1$ independent phases must be determined, full quantum state
tomography is unnecessary.
The reconstruction cost therefore scales linearly with $N$.

Finally, the diagonal unitary is reconstructed as
\begin{align}
    \Sigma^+
    =
    \mathrm{diag}\big(1, e^{i\phi_2}, \ldots, e^{i\phi_N}\big),
\end{align}
and $\Sigma^- = \Sigma^{+*}$.

\section{A demonstrative application}
\label{sec:results}

\subsection{Model System}

\begin{figure}
    \centering
    \includegraphics[width=0.5\linewidth]{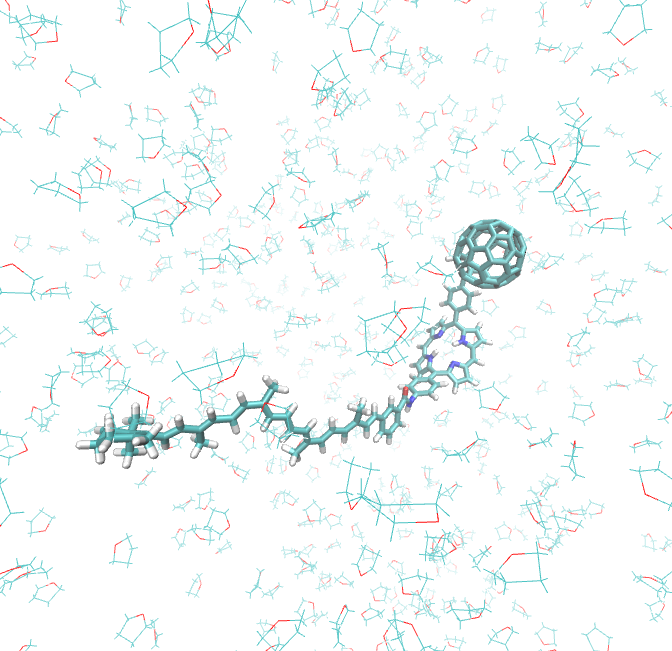}
    \caption{A snapshot of the CPC$_{60}$ molecular triad, in the bent conformation, solvated in liquid THF. }
    \label{fig:cpc60ode}
\end{figure}

In this section, we demonstrate the applicability of our quantum algorithm, which we will refer to as q-SVD-EOM, on
a model for the  $\pi\pi^* \rightarrow$ CT1  photoinduced charge
transfer dynamics in the CPC$_{60}$ molecular triad (in the bent conformation) dissolved in liquid THF at 300K (see Fig.~\ref{fig:cpc60ode}) \cite{brian_three-state_2021,tong_charge_2020,hu_photoinduced_2020,sun_computational_2018,khazaei_cavity-modified_2025}.
The model parameters were adopted from 
Ref.~\citenum{khazaei_cavity-modified_2025}. 

The nonautonomous ODE to be solved in this case is given by:
\begin{equation}
    \frac{d}{dt} 
    \left[ 
    \begin{array}{c}
    P_D (t) \\ P_A (t)
    \end{array}
    \right] 
    = \left[
    \begin{array}{cc}
      -k_{D\rightarrow A}(t) & k_{A\rightarrow D}(t) \\
      k_{D\rightarrow A}(t) & -k_{A\rightarrow D}(t)
    \end{array}
    \right]
    \left[ 
    \begin{array}{c}
    P_D (t) \\ P_A (t)
    \end{array}
    \right] 
    \label{eq:p-tcl-OD-QME}
\end{equation}
Here, $P_D (t)$ and $P_A (t)$ are the occupancies of the $\pi\pi^*$ donor state and $CT1$ acceptor state, respectively, and $k_{D \rightarrow A} (t)$ and 
$k_{A \rightarrow D} (t)$ are the donor-to-acceptor and acceptor-to-donor 
nonequilibrium Fermi's golden rule (NE-FGR) rate coefficients, which are explicitly time dependent \cite{coalson_nonequilibrium_1994,cho_nonequilibrium_1995,izmaylov_nonequilibrium_2011,sun_nonequilibrium_2016,sun_non-condon_2016,hu_photoinduced_2020,lai_simulating_2021}.   
The reader is referred to Ref.~\citenum{khazaei_cavity-modified_2025} for further discussion of this model and closed-form expressions for the NE-FGR rate coefficients.

\subsection{Quantum Simulations}

The q-SVD-EOM algorithm was implemented using Qiskit \cite{javadi-abhari_quantum_2024}
and simulated using both ideal and noisy quantum backends.
Ideal simulations employ the standard
\verb|AerSimulator|,
while noisy simulations use
\verb|AerSimulator.from_backend|
with IBM fake-device backends.

The noisy simulations employ the
\verb|FakeNighthawk| and \verb|FakePrague|
backends in order to assess the robustness
of the q-SVD-EOM under different realistic
hardware noise conditions.
The \verb|FakeNighthawk| backend corresponds
to a comparatively higher-fidelity device calibration
that is more representative of the performance of
current state-of-the-art superconducting quantum hardware,
while \verb|FakePrague| provides a more noise-sensitive
benchmark illustrating the effect of stronger hardware
imperfections on the iterative propagation.
For the present implementation,
\verb|FakeNighthawk| exhibits closer agreement with
the classical reference dynamics than
\verb|FakePrague|.

For each backend, circuits are transpiled using
Qiskit's preset pass manager with optimization
level 3, and expectation values are evaluated
using \verb|SamplerV2|.
Each circuit execution uses
$N_s = 10^6$ measurement shots.

\subsection{Results}

We simulate charge-transfer dynamics in the CPC$_{60}$ molecular
triad dissolved in THF by directly integrating Eq. \ref{eq:p-tcl-OD-QME}, with the initial condition $\{ P_D (0)=1, P_A(0)=0 \}$,
on a classical device and, independently,
via the q-SVD-EOM algorithm on a quantum simulator.
This enables a direct comparison between the classical reference
solution and q-SVD-EOM.

The total propagation time is $t_f = 10^4$ au.
The q-SVD-EOM propagation is performed over the interval
$[\tilde t,t_f]$ with $\tilde t = 50$ atomic units,
using $N_t = 400$ time steps.
Because q-SVD-EOM assumes nondegenerate singular values
(the coefficients $\{Z_{jk},K_{jk}\}$ diverge when
$\sigma_j=\sigma_k$),
the propagation cannot be initialized directly at $t=0$,
where the propagator is the identity operator and all singular values
are degenerate.
Accordingly, the full propagator $\Phi(t)$ is first obtained
classically by integrating Eq. \ref{eq:p-tcl-OD-QME} from $t=0$ to
$\tilde t = 50$au.
Three classical seed propagators,
evaluated at
$t=\tilde t-2\Delta t$,
$t=\tilde t-\Delta t$,
and
$t=\tilde t$,
are generated in order to initialize the multistep
MPEA extrapolation used in q-SVD-EOM.
A numerical SVD of these propagators
provides the initial values of
$U(t)$,
$V(t)$,
$\Sigma^\pm(t)$,
and $\sigma_1(t)$.

The subsequent SVD-EOM propagation employs the MPEA
extrapolation together with Cayley-transform updates for the
skew-symmetric generators.
The matrices $U$ and $V$ are propagated using a row-wise
decomposition, where each row vector is encoded and evolved
independently on a one-qubit quantum circuit.
The evolution of $\Sigma^\pm$ is obtained from measurements in
the $X$ and $Y$ bases, from which the relative phase is reconstructed.
The donor population is reconstructed from the SVD factors as
\begin{align}
\hat{\rho}(t)
=
\frac{\sigma_1(t)}{2}
\, U(t)
\bigl(\Sigma^+(t) + \Sigma^-(t)\bigr)
V^\dagger(t)
\, \hat{\rho}(0).
\end{align}

\begin{figure}
    \centering
    \includegraphics[width=\linewidth]{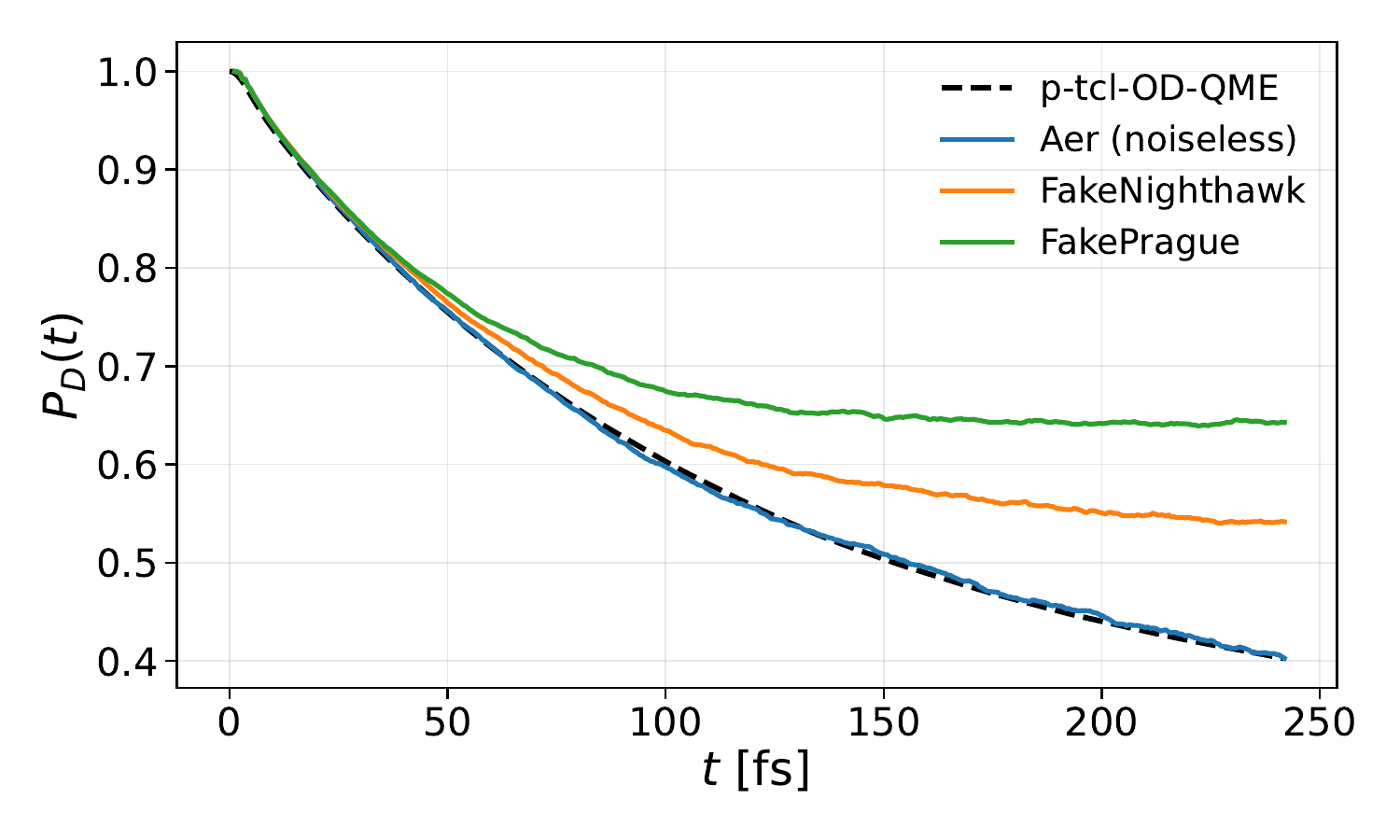}
\caption{
Comparison of $P_D (t)$ as obtained by integrating Eq. \ref{eq:p-tcl-OD-QME} on a classical computer via the second-order Runge-Kutta (RK2) method  
with $P_D (t)$ as obtained via the q-SVD-EOM algorithm on the noise-free ideal Aer simulator, as well as via noisy simulators. 
}
\label{fig:eom_comp_noise}
\end{figure}

Figure~\ref{fig:eom_comp_noise} compares $P_D (t)$ as obtained by integrating Eq. \ref{eq:p-tcl-OD-QME}
 on a classical computer by using the second-order Runge-Kutta (RK2) \cite{hairer_solving_1993} 
with $P_D (t)$ as obtained via the q-SVD-EOM algorithm on the noise-free ideal Aer simulator, as well as via noisy simulators based on
the \verb|FakeNighthawk| and \verb|FakePrague|
backends.

The ideal Aer simulation reproduces the classical
reference dynamics, demonstrating the accuracy of q-SVD-EOM in the absence of hardware noise.
Among the noisy simulations, the \verb|FakeNighthawk| backend
maintains closer agreement with the classical trajectory than
\verb|FakePrague|, reflecting differences in backend-dependent
gate and measurement fidelities.

At each time step, the quantum device implements Cayley-transform
updates for the row vectors of $U$ and $V$ as well as for the
diagonal evolution of $\Sigma^\pm$.
Because the reconstructed quantities
$\tilde{U}(t_{i+1})$,
$\tilde{V}(t_{i+1})$,
and
$\tilde{\Sigma}^{\pm}(t_{i+1})$
are used to construct the generators at subsequent time steps
through the MPEA scheme, errors propagate forward and accumulate
throughout the evolution.
Although skew-symmetry is enforced at the generator level and
optional orthogonal projection may be applied, these corrections
do not fully remove accumulated noise.
As a result, small per-step inaccuracies lead to increasingly
pronounced deviations at longer times.
This cumulative error growth reflects the sensitivity of the
iterative evolution to coherent gate errors, decoherence,
and readout noise.

Nevertheless, q-SVD-EOM remains numerically stable
throughout the full simulated time interval despite the iterative
reuse of noisy reconstructed SVD factors.
No numerical instabilities were observed during the propagation,
indicating that the structure-preserving formulation remains
numerically stable throughout the simulated time interval even
in the presence of backend noise.

The dominant computational cost arises from repeated circuit
executions and measurement sampling during the iterative evolution.
At each time step, the propagation of $U$ and $V$ requires
$N$ independent state evolutions of dimension $N$ together with
reconstruction of the corresponding row vectors from measurement
outcomes.
Further improvements could be achieved through error mitigation or
error correction techniques, as well as reductions in circuit depth.
For example, Walsh--Gray synthesis of diagonal unitaries may reduce
gate counts \cite{welch_efficient_2014,seneviratne_exact_2024}, while alternative encodings could lower the effective
dimension of the propagated states.


\section{Summary and Concluding remarks}
\label{sec:summary}

In this paper, we introduced  
a quantum algorithm, q-SVD-EOM, for simulating nonunitary dynamics governed by
nonautonomous linear ODEs. 
Rather than calculating the propagator and performing SVD on it on a classical computer at each and every time step, 
we propagate the SVD factors on the quantum computer 
through Cayley-transform
updates of the skew-symmetric generators.
Within q-SVD-EOM, the only task assigned to a classical computer is the propagation of the largest singular value,
thus making the algorithm almost fully quantum, with minimal classical computational overhead. 


The applicability of the quantum algorithm was demonstrated by using it to simulate the 
photoinduced charge transfer dynamics in the
CPC$_{60}$ molecular triad dissolved in THF.
The SVD-EOM propagation was implemented using the MPEA
extrapolation together with a row-wise decomposition of the
unitary matrices $U$ and $V$, where each row vector is propagated
independently on a one-qubit quantum circuit rather than through
a full Liouville-space vectorization of the matrices.
The diagonal matrices $\Sigma^\pm$ were propagated using
Cayley-transform updates, with their relative phases obtained
from measurements in the $X$ and $Y$ bases.
Comparison with direct classical integration demonstrates
that, in the absence of hardware noise, the q-SVD-EOM accurately reproduces the reference dynamics.

To assess the effect of realistic hardware imperfections,
we also performed noisy simulations using backend-based noise models
constructed from the IBM fake-device backends
\verb|FakeNighthawk| and \verb|FakePrague|.
The results show that the accuracy of the iterative propagation
is strongly influenced by backend-dependent gate and measurement
fidelities.
Among the noisy simulations considered here,
\verb|FakeNighthawk| exhibits closer agreement with the classical
reference dynamics than \verb|FakePrague|.

When realistic backend noise is included, deviations from the
classical reference increase over time due to accumulated gate,
decoherence, and readout errors during the iterative propagation.
Nevertheless, the results demonstrate that  q-SVD-EOM 
can reproduce the charge transfer dynamics with reasonable accuracy on
current noisy quantum simulation backends, indicating that the
approach is compatible with near-term quantum hardware.
Although skew-symmetry is enforced at the generator level and
optional orthogonal projection may be applied, these corrections
do not fully eliminate accumulated noise.

The dominant computational cost originates from repeated circuit
executions and measurement sampling required during the iterative
time evolution.
Although the present implementation is limited by circuit depth
and hardware noise, the results indicate that improvements in
gate fidelity, coherence time, and circuit optimization would
directly extend the accessible time scales and system sizes.
Additional efficiency gains may also be obtained through improved
synthesis of structured unitaries, alternative encodings, and
more efficient measurement strategies.

More broadly, q-SVD-EOM provides a general strategy
for simulating nonunitary dynamics on quantum devices by converting
the problem into coupled unitary evolutions.
Because any complex-valued linear ODE can be recast in real form,
the method can 
be extended to other time-dependent linear dynamical systems in physics and chemistry.
This work therefore establishes a proof of principle for an almost fully quantum simulation of open-system dynamics and provides a foundation
for future studies of larger and more complex nonautonomous systems.

\section{Acknowledgments}
\begin{acknowledgments}
E.G. acknowledges support from the NSF via Grant CHE-2154114.
We thank Prof. Xiang Sun for sharing the parameters for the CPC$_{60}$ harmonic three-state model Hamiltonian. The top panel of Fig. \ref{fig:cpc60ode} was created with VMD (VMD was developed by the Theoretical and Computational Biophysics Group in the Beckman Institute for Advanced Science and Technology at the University of Illinois at Urbana-Champaign \cite{humphrey_vmd_1996}). 
\end{acknowledgments}

\bibliography{svd,bib}

\end{document}